\documentclass[a4paper, notoc,floats,nofootinbib,eqsecnum,showpacs,amsmath]{JHEP3}
\usepackage[dvips]{graphicx}
\usepackage{amsfonts}
\usepackage{amssymb}
\usepackage{epsfig}
\usepackage{cite}

\usepackage{amsfonts}
\usepackage{amsmath}
\usepackage{amssymb}

\usepackage{epsfig,graphicx,amssymb,amsfonts}

\def \beq {\begin{equation}}
\def \eeq {\end{equation}}
\def \bea {\begin{eqnarray}}
\def \eea {\end{eqnarray}}

 \def\e{{\rm e}}

\def\Z#1{_{\lower2pt\hbox{$\scriptstyle#1$}}}
\def\X#1{_{\lower2pt\hbox{$\scriptscriptstyle#1$}}}

\title{Warped compactification to de Sitter space}

\author{Ishwaree P. Neupane\\
Department of Physics and Astronomy, University of Canterbury\\
Private Bag 4800, Christchurch 8041, New Zealand\\
E-mail: {\sl ishwaree.neupane@canterbury.ac.nz}}

\abstract{We explore in detail the prospects of obtaining a
four-dimensional de Sitter universe in classical supergravity
models with warped and time-independent extra dimensions,
presenting explicit cosmological solutions of the
$(4+n)$-dimensional Einstein equations with and without a bulk
cosmological constant term. For the first time in the literature
we show that there may exist a large class of warped supergravity
models with a noncompact extra dimension which lead to a finite 4D
Newton constant as well as a massless 4D graviton localized on an
inflating four-dimensional FLRW universe. This result helps
establish that the `no-go' theorem forbidding acceleration in
`standard' compactification of string/M-theory on physically
compact spaces should not apply to a general class of warped
supergravity models that allows at least one noncompact direction.
We present solutions for which the size of the radial dimension
takes a constant value in the large volume limit, providing an
explicit example of spontaneous compactification.}

\keywords{De Sitter universe; Warped extra dimensions;
Accelerating solutions}

\preprint{UOC-TP 013/10, ~ arXiv:1011.5007}

\begin{document}

\section{Introduction}

Since Kaluza's pioneering work in the early 1920s~\cite{Kaluza}
there have been a wide range of speculations about a fifth
dimension and also higher dimensions of space. In order to avoid
gross violations with everyday experience and numerous tabletop
and collider experiments, it is thought that extra dimensions
would have to be compact and tiny, or curled-up very tightly so
that they have an extremely small radius: $10^{-30}$ cm or less,
and also that their effects on present day experiments become
unobservably small. However, this may not be necessarily the case
if the background (spacetime) geometry is non-factorizable or
warped,  and the theory allows at least one noncompact extra
dimension~\cite{RS1,RS2,Binetruy:1999,Neupane:2001}. Recent
studies, see, e.g.~\cite{Shiromizu99,Garriga99,Maartens:2010},
show that cosmological models with a noncompact extra space are
consistent with known physics, and can help solve some vexing
problems in particle physics and cosmology, including the mass
hierarchy and cosmological constant problems.

The possibility of constructing four-dimensional matter-gauge
theories from higher-dimensional theories of
gravity~\cite{Polchinski95} has led to a variety of viable models
of the early universe cosmology, including the possibility of
explaining cosmic inflation using branes (or brane-antibrane
pairs)~\cite{Dvali:1998}. This idea has inspired many theoretical
physicists to explore and investigate models of de Sitter
cosmology in the context of string/M-theory as well as in simplest
braneworld models~\cite{Langlois00a,Langlois03a,Sasaki05a}. The
KKLT model~\cite{KKLT} is an example `flux compactification' in
string theory - roughly, a model building in which certain
background fluxes are turned on and non-perturbative effects are
invoked for the purpose of stabilizing the extra-dimensional
volume, which is otherwise left unfixed within the KKLT approach.

It is worthwhile to realize a state of de Sitter expansion in four
dimensions as an explicit solution to Einstein field equations in
higher dimensions, first without introducing any stringy
corrections to Einstein gravity. This seems to be difficult from
the viewpoint of the `no-go" theorem discussed by
Gibbons~\cite{Gibbons-84}, De Wit et al.~\cite{WSH87}, Maldacena
and Nunez~\cite{Malda-Nunez} and many others~\cite{GKP}. The
theorem basically asserts that if we dimensionally compactify any
string-derived supergravity models on a smooth compact manifold
${\cal M}$, then we find a flat Minkowski or anti de Sitter
spacetime as a background solution of classical supergravities in
$4+n$ dimensions unless that we also violate certain positivity
conditions.

In recent years, many authors have constructed varieties of
time-dependent solutions. One of the motivations for this has been
that the original no-go theorem required time independence of the
internal space, so one could look for time-dependent solutions in
higher
dimensions~\cite{TW03,Ohta03,Ish05,Ish03,Maeda-all,Ish03a,Neupane06}.
Through many examples discussed in the literature, we have learnt
that higher-dimensional theories with time-dependent metric moduli
generally give rise to a transient acceleration (of the universe).
Moreover, cosmological solutions with time-dependent metric
scalars possess some kind of metric singularities, especially,
when the extra-dimensional manifold is hyperbolic or contains a
subspace that is negatively curved. Additionally, cosmological
solutions obtained by allowing time-dependent metric moduli
possess a particular drawback that some of the `fundamental
constants' in nature, such as, Yukawa couplings, can vary with
time, when the model is coupled with matter fields. This result is
not encouraging since experimental and astrophysical bounds
applicable to such variations place strong constraints on (or even
rule out) these models.

In this paper we find interest in models of warped
compactifications for which the extra dimensions are
time-independent. As a viable alternative to standard Kaluza-Klein
type compactifications with physically compact extra dimensions,
we choose to `compactify' a class of string-inspired supergravity
models in $4+n$ dimensions by allowing a non-compact dimension
plus a $(n-1)$-dimensional compact manifold. Here we focus our
discussions on a class of explicit cosmological solutions which
give rise to a positive cosmological constant in the usual four
dimensions. We explore de Sitter solutions within a general class
of warped supergravity for which the 4D Newton constant is finite
and the 4D massless graviton wavefunction is normalizable in an
inflating brane or FLRW universe. This result is new and quite
remarkable. The original no-go
theorems~\cite{Gibbons-84,WSH87,Malda-Nunez} are about
(physically) compact internal spaces and their associated
subtleties, whereas the model studied here has a noncompact
direction. Our solutions nevertheless give a finite 4D Newton
constant and a concrete realization of four-dimensional Einstein
gravity on a de Sitter brane.

We also show that cosmological models with an inflating
Friedmann-Robertson-Walker universe embedded in a five-dimensional
de Sitter space give rise to a finite 4D Planck mass similar to
that in the Randall-Sundrum type braneworld models in static
AdS$_5$ spacetimes. In spacetime dimensions $D\equiv 4+n \ge 7$,
however, we find that a negative bulk cosmological term may be
preferred over a positive cosmological term.

\section{De Sitter solutions in various dimensions}

We find interest in $(4+n)$-dimensional warped metrics that
maintain the symmetries of a Robertson-Walker cosmology in four
dimensions:
\begin{equation}\label{10d-metric-gen}
ds_{D}^2 = e^{2 A(y)}\, \hat{g}_{\mu\nu} dx^\mu dx^\nu +  e^{2
B(y)} {g}_{mn}(y)\, dy^m dy^n,
\end{equation}
where $x^\mu$ are the usual spacetime coordinates ($\mu, \nu =0,
1, 2, 3$) and, $A(y)$ and $B(y)$ are some functions of one of the
internal coordinates, $y$. The internal space metric i.e.
$ds_n^2={g}_{mn}(y) dy^m dy^n$ may be taken to be a general
Einstein space, not just a constant curvature manifold.

\subsection{The $D=5$ case}

The basic idea behind the existence of a four-dimensional de
Sitter space solution (${\rm dS}_4$) supported by warping of extra
spaces can be illustrated by considering a five-dimensional
`warped metric',
\begin{equation}
ds_{5}^2 = e^{2 A(y)}\, \hat{g}_{\mu\nu} dx^\mu dx^\nu +  e^{2
B(y)} \rho^2\, dy^2, \label{5d-main-anz}
\end{equation}
where $\rho$ is a compactification radius, and the 4D metric in
the standard Friedmann-Lema\^itre-Robertson-Walker (FLRW) form
\begin{equation}\label{FRW}
ds_4^2 \equiv \hat{g}_{\mu\nu} dx^\mu dx^\nu =   -dt^2+
a^2(t)\left[\frac{ dX^2}{1-k X^2}+ X^2 (dY^2+\sin^2Y dZ^2)\right].
\end{equation}
Here the 3D curvature constant $k$ is arbitrary~\footnote{ Of
course, one could choose to rescale $k$ to $0$ or $\pm 1$, but in
this case one would also have to rescale all other dimensionful
quantities. There is no need to fix $k$ in a particular way as
long as one can solve the $D$-dimensional Einstein equations with
an arbitrary $k$.}. It is not difficult to check that the metric
Ansatz (\ref{5d-main-anz}) becomes an exact solution of 5D
Einstein equations, following from
\begin{equation}
S \propto   \int d^5{x} \sqrt{-g} R,
\end{equation}
when
\begin{equation}
B(y)=A(y) + \ln \frac{dA}{dy}-\ln c\Z{1}, \quad a(t)= \frac{1}{2}
\left(c\Z{0} \, \exp(c\Z{1} t/\rho) + \frac{k \rho^2}{c\Z{0}}
\exp(-c\Z{1} t/\rho)\right).\label{sol-5d-1}
\end{equation}
The integration constant $c\Z{1}$ may be set to unity or absorbed
in $\rho$. Further, $c\Z{0}$ may also be set to unity by using the
freedom to rescale time, i.e. $t\to t-t\Z{0}$. This result shows
that cosmologies with de Sitter expansion for late times can be
produced with different choices of $B(y)$. In the particular case
that $B(y)=0$, the solution is given by
\begin{equation}
e^{2A} = \frac{(y-y_0)^2}{y_1^2}.
\end{equation}
Since the warp factor vanishes at $y=y_0$, this solution is not
particularly interesting.

To obtain a completely regular solution, we have to make a
suitable choice of the warp factor. For $A(y)=B(y)$, we can write
the 5D metric as
\begin{equation}\label{5d-anz-2}
ds_{D}^2 = e^{2 A(y)} \left( \hat{g}_{\mu\nu} dx^\mu dx^\nu +
\rho^2 \beta^2\,dy^2\right).
\end{equation}
The numerical constant $\beta$ has been introduced just for
convenience. The 5D Einstein equations are now explicitly solved
when
\begin{equation}
e^{2A(y)} =  e^{- 2 \beta (y+y_0)}
\end{equation}
with the 4D scale factor as given above (cf~Eq.(\ref{sol-5d-1})).
The slope of the warp factor is determined in terms of the
constant $\beta$. In the above case, the warp factor diverges as
$y\to -\infty$. As a result, the warped volume is not finite
unless that one imposes a $Z_2$ symmetry along the $y=0$
hypersurface or as in RS models.

To this end, we may introduce a bulk cosmological term into the 5D
action
\begin{equation}
S = M\Z{(5)}^3   \int d^5{x} \sqrt{-g} \left(R -2\Lambda_5\right),
\end{equation}
where $\Lambda_5$ is the 5D bulk cosmological constant. With the
metric Ansatz~(\ref{5d-anz-2}), the 5D Einstein equation are
explicitly solved when
\begin{equation}
A(y)= \ln \left(\frac{b^2}{\rho^2}\right)- \frac{1}{2}\ln
\Big(\exp(\beta y)+\frac{\Lambda_5 b^4}{24\rho^2}\,\exp(- \beta
y)\Big)^2,
\end{equation} where $b$ is a
constant with length dimension of one, and the 4D scale factor is
given by
\begin{equation}
a(t)= \frac{c\Z{0}^2+k\rho^2}{2c\Z{0}}\,\cosh (t/\rho) +
\frac{c\Z{0}^2-k\rho^2}{2c\Z{0}}\,\sinh (t/\rho),
\end{equation}
where $c\Z{0}$ is an integration constant. The constant $k$
determines the asymptotic form of $a(t)$ in the usual way. In a
spatially flat universe ($k=0$), the scale factor grows as
$a(t)\propto e^{(t/\rho)}$. In a non-flat universe ($k\ne 0$), the
expansion is close to being de Sitter. In spacetime regions
endowed with a strong gravitational potential, the curvature $k$
is always non-vanishing, at least, in a local region. When one
considers an infinitely large universe, with the Hubble radius $r
\equiv c H_0^{-1} \sim 10^{27}~{\rm cm}$, then $a(t)\simeq a_0
\,e^{t/\rho}$. In fact, in the $k=0$ case, with $a(t) \propto e^{H
t}$, the solution for the warp factor can be written in a slightly
more general form
\begin{equation}
A(y)=\ln (b H) - \frac{1}{2} \left( e^{\rho H y} + \frac{
\Lambda_5 b^2}{24}\,e^{-\rho H y}\right)^2,
\end{equation}
where $\rho$ and $H$ are arbitrary constants.

Though the radial coordinate $y$ is non-compact, because of a
significant warping of the fifth dimension, we can get a finite 1d
warped volume, especially, with $0<\lambda<1$. To be more precise,
from the explicit solution given above above, we
derive~\cite{Ish09a,Ish09b})
\begin{eqnarray}
M_{(5)}^3 \int d^5{x} \sqrt{g}\, R &=& M_5^3 \, \beta\rho  \int
e^{3A(y)} dy \int \sqrt{-\hat{g}_4} \left(\hat{R}_4-\frac{12
{A^\prime}^2}{\beta^2\rho^2} -
\frac{8A^{\prime\prime}}{\beta^2\rho^2}-2\Lambda_5
e^{2A}\right)\nonumber
\\
&=&M_{\rm Pl}^2 \int d^4{x} \sqrt{-\hat{g}_4}  \, \hat{R}_4 -
M_{(5)}^3
 \,\frac{\beta\,b^6}{\rho^5} \int d^4{x}
\sqrt{-\hat{g}_4} \int dy\, \Lambda(y),\nonumber \\
\end{eqnarray}
where
\begin{equation}
M_{\rm Pl}^2 = M_{(5)}^3\, \beta \frac{b^6}{\rho^5}
\int_{-\infty}^{\infty} \frac{dy}{\left( \e^{\beta y} + \lambda \,
\e^{-\beta y}\right)^{3}}, \quad \Lambda(y) \equiv \frac{4 \left(
3\,e^{2 \beta y} +3\lambda^2 \,e^{-2\beta y} -2\lambda
\right)}{\rho^2 \left( e^{\beta y}+ \lambda \,e^{-\beta y}
\right)^5}.
\end{equation}

\subsubsection{The $\lambda >0$ case} In the $\lambda>0$ (and
hence $\Lambda_5>0$) case, we explicitly derive
\begin{eqnarray}
M_{\rm Pl}^2 =  M_{(5)}^3\,\frac{b^6}{\rho^5}\,
\frac{\tan^{-1}\sqrt{\lambda}+\cot^{-1}\sqrt{\lambda}}{8
{\lambda}^{3/2}}.
\end{eqnarray}
Since
\begin{equation}
\tanh^{-1}\sqrt{\lambda}+ \cot^{-1}\sqrt{\lambda}=\frac{\pi}{2},
\quad \Lambda_5 \equiv \frac{6}{\ell_{\rm dS}^2}, \quad
\lambda=\frac{\Lambda_5 b^4}{24\rho^2},
\end{equation}
the 4D effective Planck mass is simply given by
\begin{equation}
M_{\rm Pl}^2 = \frac{\pi}{2} M_{(5)}^3  \frac{1}{\rho^2}
\left(\frac{6}{\Lambda_5}\right)^{3/2}.
\end{equation}
The $y$ coordinate can vary from $-\infty$ to $+\infty$ and the 5D
spacetime is geodesically complete.

\subsubsection{The $\lambda < 0$ case}

In this case, the effective 4D Planck mass diverges when $y$ is
integrated from $-\infty$ to $+\infty$, independent of the sign of
the coefficient $\beta$. To get a finite 4D Planck mass, $M_{\rm
Pl}$, one may allow the $y$ coordinate to range from $0$ to
$+\infty$ (for $\beta>0$) or from $-\infty$ to $0$ (for
$\beta<0$). This is similar to that in Randall-Sundrum braneworld
models, where one considers only half of the AdS$_5$ space and
replace the other half by its mirror image.

\subsubsection{The effect of brane action}

If required, one may supplement the 5D gravity action with brane
action
\begin{equation}
S_{\rm brane}=  \int_{\partial {\cal M}_1}
\sqrt{-g_{b1}}\,(-\tau_{b1}) + \int_{\partial {\cal M}_2}
\sqrt{-g_{b2}}\,(-\tau_{b2}),
\end{equation}
where $\tau_{b1}$, $\tau_{b2}$ denote the brane tensions
corresponding to the two 3-branes $b1$ and $b2$, and then solve
the 5D Einstein equations by placing two 3-branes at $y=0$ and
$y=\pi$. In such a configuration, while computing derivatives of
$A(y)$, we have to consider the metric a periodic function in $y$.
The solution valid for $-\pi \le y\le \pi$ then implies that
\begin{equation}\label{two-deri}
A^{\prime\prime}+ \frac{\beta^2 \rho^2 b^4  \Lambda_5}{6 \,K_+^2}
+\frac{\beta K_-}{K_+} \left(2\delta (y) - 2\delta
(y-\pi)\right)=0,
\end{equation}
where
\begin{equation}
K_{\pm} \equiv e^{\beta\, y} \pm  \frac{\Lambda_5 b^4}{24
\rho^2}\, e^{- \beta\, y}.
\end{equation}
From the $\mu\nu$-components of the 5D Einstein equations we get
\begin{eqnarray}
 \frac{3}{\beta^2\rho^2} \left(A^{\prime\prime}+{A^\prime}^2 - \beta^2\right)
 + \Lambda_5 \, e^{2A} +\frac{e^{A}}{2\beta \rho M_5^3}
\left(\tau_{b1} \delta(y) + \tau_{b2}
\delta(y-\pi)\right)&=& 0\nonumber \\
\Rightarrow  A^{\prime\prime}+ \frac{\beta^2 \rho^2 b^4
\Lambda_5}{6\,K_+^2} +\frac{ \beta b^2}{6 \rho M_5^3\, K_+}
\left(\tau_{b1}\delta (y) + \tau_{b2} \delta (y-\pi)\right)&=&
0.\label{b-tensions}
\end{eqnarray}
By comparing Eqs.~(\ref{two-deri}) and (\ref{b-tensions}), we get
\begin{equation}
\tau_{b1}=\frac{3 M_5^3\,\rho}{b^2} \left(1-\lambda \right), \quad
\tau_{b2}=- \frac{3 M_5^3\,\rho}{b^2} \left(e^{\beta\, \pi }
-\lambda \,e^{-\beta \,\pi}\right),
\end{equation}
where
\begin{equation}
\lambda \equiv \frac{\Lambda_5 b^4}{24 \rho^2}.
\end{equation}
The solution found in~\cite{Sasaki2000a} corresponds to the choice
$\lambda=-1$ (so that $\Lambda\Z{5}<0$), while that
in~\cite{Aguilar10a} corresponds to $\lambda=1 $. In both these
papers $\rho$ was set to unity. In the discussion below, we will
take $\lambda>0$ and also send the second brane to infinity, or
simply replace the 3-brane with a physical four-dimensional FLRW
universe. A detailed discussion on localization of gravity in five
dimensions appears elsewhere~\cite{Ish10d}.

In the following we will not compute the brane tension explicitly,
but we note that the corresponding solutions give rise to a
positive brane tension when a p-brane with $3\le p\le (D-2)$ is
introduced along with the gravitational action. The location of
the brane may be viewed as the place where the zero mode graviton
wavefunction is peaked.

\subsection{The $D=6$ case}

With two extra dimensions, we may write the metric in the form
\begin{equation}\label{6D-ansatz}
ds_6^2 = e^{2A(y)} \, \hat{g}_{\mu\nu} dx^\mu dx^\nu + e^{2 B(y)}
\rho^2 \left( e^{2C(y)} dy^2 + e^{2D(y)} d\theta^2\right),
\end{equation}
where $\theta$ is a periodic coordinate, $0\le \theta\le 2\pi$.
The internal space becomes physically compact when $C(y)=0$ and
$D(y)=\ln\sin{y}$. But in this case the warp factor becomes
singular at $y=0$. We shall make some other choices that yield a
regular warp factor.

It is not difficult to check that the metric Ansatz
(\ref{6D-ansatz}) becomes an exact solution of 6D Einstein
equations, following from
\begin{equation}
S \propto   \int d^6{x} \sqrt{-g} R,
\end{equation}
for instance, when
\begin{equation}\label{main-sol-scale1}
e^{2A(y)} = \frac{h^2}{X^{2p}}, \quad e^{2B+ 2C}=  h^2\, \frac{p^2
{X^\prime}^2}{X^{2p+2}}, \quad e^{2B+ 2D}= h^2,
\end{equation}
where $X\equiv X(y)$, $X^\prime= dX(y)/dy$, $h$ and $p$ are
dimensionless constants, and
\begin{equation}
a(t)=\frac{c\Z{0}^2+k \rho^2}{2c\Z{0}}\,\cosh (t/\rho) +
\frac{c\Z{0}^2-k \rho^2}{2 c\Z{0}}\,\sinh
(t/\rho).\label{sol-4Dscale}
\end{equation}
The 3D spatial curvature constant has a dimension of $({\rm
length})^{-2}$. As is evident, the scale factor is regular
everywhere with $k=0$ and $k>0$. Especially, in the case $k<0$
(i.e., in an open universe), and with the choice $k
\rho^2=-c\Z{0}^2$, we get $a(t)=\sinh(\sqrt{|k|}t/c\Z{0})$.

For instance, with $X\equiv \cosh(M y)$, the six-dimensional
solution takes the form
\begin{equation}\label{sol-6D}
ds_6^2 = \frac{h^2}{X^{2p}} \left(\hat{g}_{\mu\nu}  dx^\mu dx^\nu
+ \rho^2 \left( \frac{p^2 M^2 (X^2-1)}{X^2}\, dy^2 + X^{2p}\,
d\theta^2\right)\right)
\end{equation}
In this case the internal two-dimensional space, given by the
metric
\begin{equation}
ds_2^2= \rho^2 \left( p^2 M^2 \frac{X^2-1}{X^2}\, dy^2 + X^{2p}\,
dz^2\right)
\end{equation} is negatively curved, $R\Z{2}\equiv  {R}_{mn} {g}^{mn}
=- {2}/{\rho^2}<0$. However, when we take into account the effect
of warping and write the internal 2d metric as
\begin{equation}
\widetilde{ds}_2^2 = \frac{h^2 \rho^2}{X^{2p}} \left(p^2 M^2
\frac{X^2-1}{X^2}\, dy^2 +X^{2p}\, dz^2\right),
\end{equation}
then we find that the 2d warped space is Ricci flat and its volume
is effectively finite.

To quantify this, we shall consider the following dimensionally
reduced action
\begin{eqnarray}
\int d^{6} x \sqrt{-g}\, R &=& p M \rho^2 \, h^4\int
\frac{\sqrt{X^2-1}}{X^{3p+1}} \, dy\int_0^{2\pi} d\theta
\int d^4{x} \sqrt{-\hat{g}_4} \left( \hat{R}_4-\frac{12}{\rho^2} \right)\nonumber \\
&= & 2\pi p \rho^2\,h^4 V\Z{2}^{\rm w} \int d^4{x}
\sqrt{-\hat{g}_4} \left( \hat{R}_4- 2\Lambda_4 \right),
\end{eqnarray}
where we have used the solution (\ref{sol-6D}) with $X \equiv
\cosh (M y)$ and allowed the range of integration from $y=-\infty$
to $y=+\infty$. The 2d warped volume $V\Z{2}^{\rm w}$ is given by
\begin{eqnarray}
V\Z{2}^{\rm w}= \frac{1}{3p} + \frac{\sqrt{\pi}\, \Gamma
[3p/2]}{2\Gamma [(3p+1)/2]} + \frac{2^{3p+1}}{2+3p} \,
{}_2F_1\left[\frac{3p+2}{2},3p+1,\frac{3p+4}{2},-1\right]
\end{eqnarray}
and $\Lambda_4\equiv 6/\rho^2$. Here we allow $p$ to take a
rational number such that $n\equiv 3p+1$ is a positive integer,
excluding $n= 0$ and $1$. We then get $V\Z{2}^{\rm w}={2}/{3p}$
and hence
\begin{eqnarray}
M_{(6)}^4 \int d^{6} x \sqrt{-g}\, R &=& M_{\rm Pl}^2 \int d^4{x}
\sqrt{\hat{g}_4} \left( \hat{R}_4- 2\Lambda \right),
\end{eqnarray}
where
\begin{equation}
 M_{\rm Pl}^2 = M_{(6)}^4 {\cal R}^2, \qquad {\cal R} \sim
 \rho \sqrt{\frac{4\pi}{3}} h^2,
\end{equation}
with ${\cal R}$ being the effective size of the extra dimensions.

In fact, in dimensions $D\ge 6$, we can find a more general class
of solutions than the ones presented above. In the $D=6$ case, the
solution may be written as
\begin{equation}\label{sol-6D-b}
ds_6^2 = \frac{h^2}{F(y)} \left[-dt^2+ a(t)^2 d\vec{x}_{3,k}^2 +
\rho^2 \left(G(y)\, dy^2 +  H(y)\, d\theta^2\right) \right],
\end{equation}
with the scale factor as given in (\ref{sol-4Dscale}) and
\begin{equation}
F(y)= \left(X + \sigma \sqrt{X^2-1}\right)^{2p}=H(y), \quad G(y)=
p^2 M^2 \left(\frac{\sqrt{X^2-1}+\sigma
X}{X+\sigma\sqrt{X^2-1}}\right)^2,
\end{equation}
where $X=\cosh (M y)$ and $\sigma$ is a numerical constant.

One should note that the $D=5$ case is special for which
Einstein's equations enforce one to take $\sigma=1$, which leads
to a diverging warp factor at $y=-\infty$ unless that one imposes
$Z_2$ symmetry along $y=0$ or introduces a positive bulk
cosmological term. In dimensions $D\ge 6$, however, there exist no
such restrictions, although for generality of the model, one may
introduce a bulk cosmological term and/or p-form gauge fields (or
fluxes).

\subsection{The $D=7$ case}

In spacetime dimensions $D=7$ (or $n=3$), the metric ansatz can be
written as
\begin{equation}\label{7-ansatz}
ds\Z{7}^2 = \frac{h^2}{F(y)} \bigg(\hat{g}_{\mu\nu} dx^\mu dx^\nu
+ \rho^2 \Big( G(y)\, dy^2 +  H(y) \left(d\theta^2+\sin^2\theta
d\phi^2\right)\Big)\bigg).
\end{equation}
This metric becomes an exact solution of 7D Einstein equations,
following from
\begin{equation}
S \propto   \int d^7{x} \sqrt{-g} \,R,
\end{equation}
for instance, when
\begin{equation}\label{sol-7D}
F(y)= \left(X + \sigma  \sqrt{X^2-1}\right)^{2p}, \quad G(y)=
\frac{5p^2 M^2}{3} \left(\frac{\sqrt{X^2-1}+\sigma X}{X+\sigma
\sqrt{X^2-1}}\right)^2, \quad H(y)=\frac{1}{3},
\end{equation}
where $X\equiv \cosh (M y)$. The 4D scale factor is still given by
(\ref{sol-4Dscale}). It is not difficult to check that the
internal 3-space, given by the metric
\begin{equation}
ds_3^2= \frac{\rho^2}{3} \left(
\frac{5}{4}\frac{{F^\prime}^2}{F^2} \,dy^2 + d\Omega_2^2\right),
\end{equation}
is positively curved, ${R}_3= 6/\rho^2$. When we take into account
the warp factor, i.e.,
\begin{equation}
\widetilde{ds}_3^2= \frac{h^2}{F(y)}\, ds_3^2,
\end{equation}
then also we find that the 3d curvature is positive,
$\widetilde{R}_3 = \tilde{R}_{mn} \tilde{g}^{mn} = {24
F(y)}/(5\rho^2 h^4)>0$.

From the explicit 7D solution given above, we derive
\begin{eqnarray}
\int d^{7} x \sqrt{-g}\, R &=& h^5\, \frac{2\pi \sqrt{5} p M
\rho^3}{3\sqrt{3}}   \int \frac{u^\prime\, dy}{u^{5p+1}}
\int d^4{x} \sqrt{\hat{g}_4} \left( \hat{R}_4- \frac{12}{\rho^2} \right)\nonumber \\
&= & h^5 \, 2\pi  \frac{\sqrt{5}\, \rho^3}{15\sqrt{3}}
\left(\frac{-1}{u^{5p}}\right) \int d^4{x}
\sqrt{\hat{g}_4}\,\left( \hat{R}_4-\frac{12}{\rho^2} \right),
\end{eqnarray}
where
\begin{equation}
u(y)\equiv \cosh (M y) + \sigma  |\sinh (M y)|.
\end{equation}
The function $u(y)$ is symmetric along $y=y_0$~\footnote{The
actual value of $y_0$ depends on the choice of $\sigma $. One has
$u=u_{\rm min}$ and $f=f_{\rm max}$ at $y=y_0$.} and hence
$\left[-1/u^{5p}\right]_{-\infty}^{\infty} \simeq 2$, provided
that $p> 0$ and $1+\sigma > 0$. Finally, under the dimensional
reduction from 7 to 4, we find
\begin{eqnarray}
M_{{\rm Pl} (7)}^5 \int d^{7} x \sqrt{-g}\, R &=& M_{\rm Pl}^2
\int d^4{x} \sqrt{\hat{g}_4} \left( \hat{R}_4- 2\Lambda \right),
\end{eqnarray}
where $\Lambda_4= 6/\rho^2$ and
\begin{equation}
M_{\rm Pl}^2 = M_{{\rm Pl}(7)}^5 {\cal R}^3, \qquad {\cal R} \sim
\rho\left(\frac{2 V\Z{2} \sqrt{5}\,h^5}{15\sqrt{3}}\right)^{1/3} .
\end{equation}
The numerical constant $h$ in the metric does not have to be set
to unity; its value is rather fixed by the D-dimensional dilaton
coupling and it can be much smaller than unity.

\subsection{Generalization to $D$-dimensions}

Next we shall begin with a general $(4+n)$-dimensional `warped
metric' of the form
\begin{equation}\label{4m-ansatz}
ds_D^2 = \frac{h^2}{F(y)}\bigg( \hat{g}_{\mu\nu} dx^\mu dx^\nu +
\rho^2 \left(G(y) dy^2 +E(y)\,d\Omega_{n-1}^2\right)\bigg).
\end{equation}
Here and henceforth we shall take $n\ge 3$~\footnote{To perform a
consistent truncation in dimensions $D\equiv 4+n \ge 7$, we can
write the internal $(D-4)$-dimensional space as a general warped
product of a non-compact direction and a compact Einstein space X
of dimensions $(D-5)$. In this set up, one just has to find a
physical way to limit the growth in the radial direction, leading
to a finite warped volume.}. This metric ansatz becomes an exact
solution of $D$-dimensional Einstein equations, following from
\begin{equation}
S = M_{(D)}^{D-2}  \int d^{D} {x} \sqrt{-g} \,R,
\end{equation}
when
\begin{equation}\label{sol-4m}
G(y)=\frac{n+2}{12} \left(\frac{F^\prime}{F}\right)^2, \qquad
E(y)=\frac{(n-2)}{3}={\rm const},
\end{equation}
and
\begin{equation}
a(t)=\frac{c\Z{0}^2+k\rho^2}{2 c\Z{0}}\,\cosh (t/\rho) +
\frac{c\Z{0}^2-k\rho^2}{2 c\Z{0}}\,\sinh (t/\rho).
\end{equation}
As is evident, cosmologies with de Sitter expansion for late times
can be produced with different choices of $F(y)$. In view of this
result, we may comfortably conclude that the usual `no-go' theorem
forbidding acceleration in `standard' compactification of
string/M-theory on physically compact spaces should not apply to a
general class of warped supergravity models that allows a
noncompact direction.

In the following discussion, we take~\footnote{This form of warp
factor can be supported, for instance, by a D-dimensional bulk
scalar field, with a nonzero scalar potential, see, for example,
refs.~\cite{Yamauchi07,Dzhunushaliev}.}
\begin{equation}\label{soln-4m}
F(y)=\left(X+\sigma  \sqrt{X^2-1}\right)^{2p}, \qquad X\equiv
\cosh(My),
\end{equation}
where $M$ is a constant with mass dimension and $p$ is a numerical
constant. Hence
\begin{equation}
G(y)=\frac{(n+2)p^2 M^2}{3}\, \frac{\sqrt{X^2-1}+\sigma
X}{X+\sigma \sqrt{X^2-1}}.
\end{equation}
The result reported in~\cite{Ish10b} is obtained by taking $n=6$,
and setting $M=1$. We can also obtain the result presented
in~\cite{Ish09} by taking $n=6$, $\sigma=0$ and choosing the
coefficient $p$ appropriately, but $p<0$. In this last case,
however, the six-dimensional warped volume is not finite unless
one also introduces some elements of the Randall-Sundrum type
braneworld models and/or introduces an ultraviolet (UV) cutoff in
the $y$-direction.

In the particular case that $\sigma=-1$, the function $F(y) \to 0$
as $y\to \infty$. We will actually discard this choice since in
this case the warp volume can be infinitely large. To get a
physical model, we shall take $1+\sigma>0$ and also take $p>0$.
This choice helps us to get a finite 4D Planck mass as well as a
normalizable zero mode graviton wavefunction.

From the explicit solution given above, we derive
\begin{eqnarray}
\int d^{4+n} x \sqrt{-g}\, R &=& h^{2+n}  \rho^{n}\,p M
\sqrt{\frac{(n+2) (n-2)^{(n-1)}}{3^n}}\nonumber\\
&{}& \quad \times V\Z{n-1} \times I(y) \times \int d^4{x}
\sqrt{\hat{g}_4} \left( \hat{R}_4- \frac{12}{\rho^2}
\right),\label{dim-reduc-1}
\end{eqnarray}
where $V\Z{n-1}$ is the volume of $S^{n-1}$ sphere,
$\Lambda_4\equiv 6/\rho^2$ and
\begin{equation}
I(y)\equiv \int_{0}^{\infty}
\frac{\sqrt{\cosh^2(My)-1}+\sigma\cosh (My)} {\left(\cosh (M
y)+\sigma\sqrt{\cosh^2 (M y) -1}\right)^{p(n+2)+1}}\,dy \simeq
\frac{1}{p(n+2)M}.
\end{equation}
In the second step above we have assumed that $ 0 \le  \sigma \ll
1$. Finally, upon the dimensional reduction from $D=4+n$ to $D=4$,
we get
\begin{equation}
\int d^{4+n} x \sqrt{-g}\, R=  h^{2+n} \rho^{n}\,
\sqrt{\frac{(n-2)^{(n-1)}}{(n+2)\,3^n}}\,V\Z{n-1}\int d^4{x}
\sqrt{\hat{g}_4}\,\left( \hat{R}_4- 2\Lambda_4 \right).
\end{equation}
Equivalently,
\begin{eqnarray}
M_{{\rm Pl}\,(4+n)}^{2+n} \int d^{4+n} x \sqrt{-g}\, R &= & M_{\rm
Pl}^2 \int d^4{x} \sqrt{\hat{g}_4} \left( \hat{R}_4-
2\Lambda\Z{4}\right),
\end{eqnarray}
where
\begin{eqnarray}
 M_{\rm Pl}^2 &=&  M_{{\rm Pl},(4+n)}^{n+2} \rho^n \,  V_{(n-1)}\,
 h^{n+2}
\sqrt{\frac{(n-2)^{(n-1)}}{(n+2) 3^n}}. \label{redu-formula}
\end{eqnarray}
For instance, with $n=3$ (i.e. $D=7$), this yields
\begin{equation}
 M_{\rm Pl}^2 = \frac{2\pi}{3\sqrt{15}}\, M_{{\rm Pl} (7)}^{5}\, \rho^3 h^5.
 \end{equation}
This may be further analyzed by taking $\rho\sim 10^{27}~{\rm
cm}$, especially, if one wants to tune $\Lambda_4$ to the present
value of 4D cosmological constant. One then has to allow $h$ to
take an extremely small value, $h\lesssim 10^{-21} \ll 1$, so that
the 7D Planck mass $M_{{\rm Pl} (7)}\gtrsim {\rm TeV}$. A
constraint like this becomes weaker in the presence of a bulk
cosmological term and/or (supergravity) background fluxes, and
also when one applies the model to explain the early universe
inflation. For instance, with $\rho^{-1} \sim 10^{-5} ~M_{\rm
Pl}$, we get $M_{\rm Pl}\sim 10^{-3} M_{{\rm Pl} (7)} h$.

In~\cite{GH2001}, Gibbons and Hull explored the possibility of
constructing a model of de Sitter universe within a context of 10-
and 11-dimensional supergravity, by allowing a noncompact extra
dimension. The authors, however, did not find a physical model of
4D de Sitter cosmology that leads to a finite 4D Planck mass. Here
we have achieved this goal.

\subsection{Effects of a bulk cosmological term}

The explicit solutions given above can easily be generalized so as
to include the effect of a bulk cosmological term. For generality,
here we shall keep the number of extra dimensions $n$ unspecified
(except that $n\ge 3$) and write the action as
\begin{equation}
S= M\Z{{\rm Pl} (n+4)}^{2+n} \int d^{4+n}x
\left(R-2\Lambda\Z{b}\right).\label{7DwithCC}
\end{equation}
The $(4+n)$-dimensional metric ansatz may be written in the form
\begin{equation}
ds_{4+n}^2 =\frac{h^2}{F(y)}\left( -dt^2 + a(t)^2 d\vec{x}_{3,k}^2
+ G(y) dy^2 + E(y)\, d\Omega_{n-1}^2 \right).
\end{equation}
Einstein's field equations following from the action
(\ref{7DwithCC}) are explicitly solved when
\begin{equation}
a(t)= \frac{1}{2} e^{H (t-t_0)}+ \frac{k}{2H^2} e^{-H (t-t_0)},
\end{equation}
where $t_0$ is a constant, and
\begin{equation}
G(y)= \frac{(n+2)(n+3) {F^\prime(y)}^2}{12(n+3) H^2
F(y)^2-8\Lambda_b F(y)}, \quad E(y)=\frac{(n-2)}{3H^2}={\rm
const},
\end{equation}
where $F^\prime = dF(y)/dy$. In dimensions $D\ge 7$ (or $n\ge 3$),
it looks suggestive to take a negative bulk cosmological term,
$\Lambda\Z{b}<0$, in which case $G(y)$ is regular everywhere.

In the discussion below, we simplify the model by taking $F\propto
(\cosh{M y})^{2 p}\equiv X^{2p}$ ($p>0$). In dimensions $D=7$, the
metric solution then takes the form
\begin{equation} ds_{7}^2 =\frac{h^2}{X^{2 p}}\left( -dt^2 +
a(t)^2 d\vec{x}_{3,k}^2 +
  \frac{15 p^2 M^2 (X^2-1)}{9 H^2
X^{2}-\Lambda\Z{b} X^{2(1-p)}}\, dy^2 + \frac{1}{3 H^2}\,
d\Omega_{2}^2 \right),
\end{equation}
which is regular everywhere with $ \Lambda\Z{b}  < 9 H^2 $. In the
$\Lambda\Z{b}<0$ case, the ratio $(-\Lambda\Z{b})/9H^2$ can take
any value between $0_+$ and $\infty$. In either case, the size of
the fifth dimension stabilizes in the large $y$ limit, implying
that the model has a feature of spontaneous stabilization.

The difference we now have as compared to the result
(\ref{dim-reduc-1}) is that $\rho=1/H$ and
\begin{equation}
I(y)=\int_{y\Z{0}}^{y\Z{\infty}}
\frac{\sqrt{X^2-1}}{X^{(n+1)p+1}\,\sqrt{X^{2p}+{\cal B}}}\, dy,
\qquad {\cal B}\equiv  \frac{|- 2\Lambda\Z{b}|}{3(n+3) H^2},
\end{equation}
where $X=\cosh(M y)$. For brevity, let us make the simplest choice
that $p=1$. Hence
\begin{equation}
I(y)=\frac{1}{16 M {\cal B}^{5/2}}\left[ 2 {\cal B} Y {\rm sech}^2
(My) \left(3-2{\cal B}\, {\rm sech}^2 (My) \right)-6 \tanh^{-1}
Y\right]_{y\Z{0}}^{y_\infty},
\end{equation}
where $Y\equiv \sqrt{(1+2{\cal B}+\cosh (My))/2{\cal B}}$. Without
loss of generality, one may take $y\Z{0}=0$ and
$y\Z{\infty}=\infty$, and consider two specific cases: ${\cal
B}\ll 1$ and ${\cal B}\gg 1$. In fact, the condition $|{\cal
B}|\ll 1$ is a reasonable approximation in the early universe,
while, in the present universe, or in a large cosmological scale,
e.g. $c H^{-1}\sim 10^{27}~{\rm cm}$, one can have ${\cal B}\gg
1$~\footnote{In terms of the bulk curvature radius $\ell \equiv
\sqrt{42/(-\Lambda\Z{b})}$, the condition ${\cal B}\gg 1$ implies
$c H^{-1}\gg \ell $.} and hence
\begin{equation} I \simeq \frac{H}{(\Lambda\Z{b})^{1/2}\,M}, \qquad
M_{\rm Pl}^2\simeq \frac{h^5 M_{{\rm Pl}, 7}^5}{H^2
\sqrt{|\Lambda\Z{b}|}}\times \frac{2\pi\sqrt{5}}{3\sqrt{3}}.
\end{equation}
That is, in a cosmological background, unlike in the simplest
braneworld models in a static AdS$_5$ background, the 4D Plank
mass may well depend on the observed value of the Hubble parameter
$H$ or the 4D cosmological constant, other than on the fundamental
$(4+n)$-dimensional Planck mass $M_{{\rm Pl}, 4+n}$ and the bulk
cosmological term $\Lambda\Z{b}$.

\section{Linearized gravity}

Higher-dimensional theories with one or more non-compact extra
dimensions require the trapping of gravitational degrees of
freedom on the brane, or a physical 4D universe. To determine
whether the spectrum of linearized tensor fluctuations $\delta
g_{AB}$ is consistent with four-dimensional experimental gravity,
we shall consider the perturbations around the background
solutions given above~\footnote{We refer to~\cite{Koyama:04a} for
the analysis of metric perturbations in higher dimensions with a
nontrivial warp factor.}. For brevity, we take $k=0$ so that the
usual 3D space is spatially flat, and hence $a(t)\propto e^{Ht}$.

In a general D dimensional spacetime (with $D=n+4 \ge 7$), the
transverse-traceless tensor modes $h_{ij}\equiv \delta
g_{ij}=\delta_i^A \delta_j^B h\Z{AB}(x^\lambda,y)$ satisfy the
following wave equation
\begin{equation}
\left( \partial_y^2 -\frac{(n+2)F^\prime}{2F} \partial_y -
\frac{G^\prime}{2 G} \partial_y - G(y) \left(\partial_t^2 +3
\frac{\dot{a}}{a}
\partial_t - a^{-2}\, \vec{\nabla}^2\right)\right)
h_{ij} =0. \end{equation} In the presence of a brane source or a
localized object at $y=y\Z{0}$, there can appear delta function
term(s) on the right-hand side of this equation. These are easily
computed once the brane action is known, but their explicit forms
are not important for our analysis.

We begin with a canonical model with $n=3$ (or $D=7$), for which
\begin{eqnarray}
\left( \partial_y^2 -\frac{5F^\prime}{2F} \partial_y -
\frac{G^\prime}{2G} \partial_y - G(y) \left(\partial_t^2 +3 H
\partial_t - e^{2Ht} \vec{\nabla}^2\right)\right)
h_{ij} =0.
\end{eqnarray}
Indeed, the $n=3$ is the case with a minimal number of extra
dimensions that allows the two-dimensional compact manifold to
have positive, negative or zero curvature. For simplicity, we have
taken the 2-dimensional compact space to be a usual two-sphere,
$S^2$. By separating the variables as
\begin{equation}
h_{ij} ( x^\mu, y) \equiv \sum \alpha_m(t) \psi_m(y)\,e^{i k\cdot
x}\,\hat{e}_{ij},
\end{equation}
where $e_{ij} (x^i)$ is a transverse, tracefree harmonics on the
spatially flat 3-space, $\vec{\nabla}^2 \hat{e}_{ij} = - k^2
\hat{e}_{ij}$, we get
\begin{eqnarray}
&&\frac{d^2\psi_m}{dy^2} - \left(\frac{5F^\prime}{2F}+
\frac{G^\prime}{2G}\right) \frac{d\psi_m}{dy} + m^2 \psi_m =
0,\label{TT-tensor}\\
&&\ddot{\alpha}_m+ 3\frac{\dot{a}}{a} \dot{\alpha}_m +
\left(\frac{k^2}{a^2}+ m^2\right) \alpha_m = 0,
\end{eqnarray}
where $m$ is a 4D mass parameter and $k$ is the co-moving
wavenumber along the 4D hypersurface.

\subsection{Tensor perturbation equation with $\Lambda\Z{b}=0$}

In the $\Lambda\Z{b}=0$ case, the background equations are
explicitly solved for
\begin{equation}
G(y)=\frac{15 {F^\prime}^2}{36 H^2 F^2} \qquad (D=7).
\end{equation}
The off-brane wave equation for tensor fluctuations, i.e.
(\ref{TT-tensor}), reduces to
\begin{equation}
\frac{d^2\psi_m}{dy^2} - \left(\frac{3F^\prime}{2F}+
\frac{F''}{F^\prime}\right) \frac{d\psi_m}{dy} + m^2
\psi_m=0.\label{7D-linear-2}
\end{equation}
By defining \begin{equation} \psi\Z{m}(y) \propto
F(y)^{3/4}\,\sqrt{\frac{dF(y)}{dy}}\, \Psi\Z{m}(y),
\end{equation}
we can bring Eq.~(\ref{7D-linear-2}) into the standard
Schr\"odinger form
\begin{equation}
\frac{d^2 \Psi\Z{m}}{dy^2}-V(y)\, \Psi\Z{m} = -m^2
\Psi\Z{m},\label{Schro-main1}
\end{equation}
where
\begin{equation}
V(y)=\frac{21}{16} \left(\frac{F^\prime}{F}\right)^2 + \frac{3}{4}
\left(\frac{F^{\prime\prime}}{F^\prime}\right)^2 - \frac{1}{2}
\frac{F^{\prime\prime\prime}}{F^\prime}.
\end{equation}
Below we consider a few explicit examples for which $F(y)$ is
regular everywhere.

\noindent Take
\begin{equation}
F(y)= \left(\cosh(M y)\right)^{2p}\equiv X^{2p}
\end{equation}
with $p>0$. The off-brane graviton wave equation takes the form
\begin{equation}
\frac{d^2\psi_m}{dy^2}-M\left((5p-1)\frac{\sqrt{X^2-1}}{X} +
\frac{X}{\sqrt{X^2-1}} \right) \frac{d\psi_m}{dy}+m^2
\psi_m=0.\label{7D-linear}
\end{equation}
By defining \begin{equation} \psi\Z{m}(y) =
\left(\frac{X^2-1}{X^2}\right)^{1/4} \,X^{5p/2}\, \Psi\Z{m}(y),
\end{equation}
we can bring Eq.~(\ref{7D-linear}) into the standard Schr\"odinger
form (\ref{Schro-main1}) with
\begin{equation}
V(y) = \frac{25p^2 M^2}{4} \left( \frac{\cosh^2(My)}{\sinh^2(My)}
+ \frac{4}{\sinh^2(2My)} - \frac{2}{\sinh^2(My)} \right)
+\frac{M^2}{\sinh^2(My)}
\left(1-\frac{1}{4\cosh^2(My)}\right).\label{potential7D}
\end{equation}
\begin{figure}[!ht]
\centerline{\includegraphics[width=4.0in,height=2.3in]
{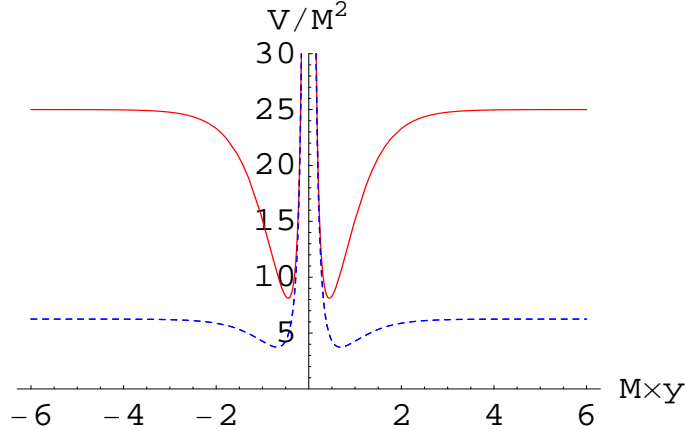}} \caption{The plot of $V(y)/M^2$ with $p=2$ (red,
solid line) and $p=1$ (blue, dotted line). } \label{poten}
\end{figure}
\begin{figure}[!ht]
\centerline{\includegraphics[width=4.0in,height=2.3in]
{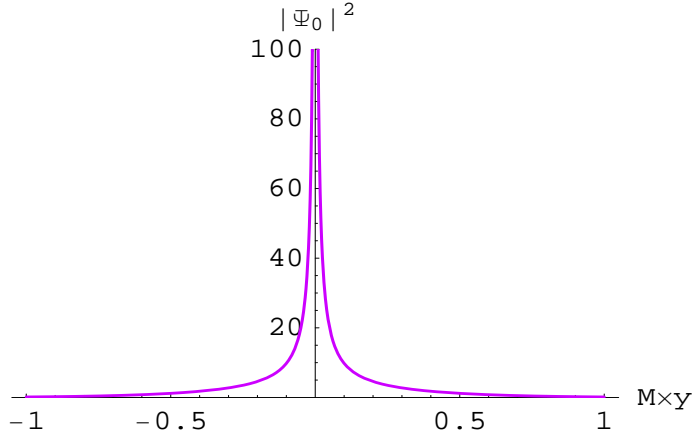}} \caption{The plot of the function $
|\Psi_0|^2$ with $N\Z{0}=1$ and $p=1$. } \label{zero-mode}
\end{figure}
Note that $V(y)\to 25p^2 M^2/4$ as $My \to \infty $ (or simply
when $M y\gg 1$) (cf Fig.~\ref{poten}). There is a mass gap, which
depends on the size of fifth dimension via $p$ and also on the
mass parameter $M$. The massless zero mode wave function is given
by
\begin{equation}
\Psi\Z{0}(y)= c\Z{1}\, \frac{\left(\cosh(M
y)\right)^{(1-5p)/2}}{\left(\cosh^2(M y)-1\right)^{1/4}}.,
\end{equation}
which is normalizable with $p>0$ and $0\lesssim \epsilon < |M y| <
\infty$ (cf Fig.~\ref{zero-mode}). Of course, in the presence of a
$\delta$-function type brane source at $y=y\Z{c}$, there would
arise an addition term in (\ref{potential7D}) proportional to
$\delta(y-y_c)$. In such a case we need satisfy the following
boundary condition
\begin{equation}
\partial_y \Psi\Z{0}(y\Z{c})=-\left(\frac{3}{4} \frac{F^\prime}{F}+
\frac{1}{2}
\frac{F^{\prime\prime}}{F^\prime}\right)\Psi\Z{0}(y\Z{c}).
\end{equation}

The precise discrete modes are given in terms of the
hypergeometric functions, and are a linear combination of
$$ X^{c_-}\left(X^2-1\right)^{3/4} \, {}_2F_1 \left(1+b_+, 1+b_-,
c_+, X^2\right)$$ and
$$X^{c_+} \left(X^2-1\right)^{3/4} \,{}_2F_1 \left(1-b_-, 1-b_+,
c_-, X^2\right),$$ where
$$ b_\pm =\frac{5p}{4} \pm
\sqrt{\frac{25p^2}{16}-\frac{m^2}{4M^2}}, \quad c_\pm = \frac{2\pm
5p}{2}, \quad X=\cosh(M y).$$ Note that, unlike in five
dimensions, the mass of KK modes is determined in terms of an
arbitrary parameter $M$ (rather than the Hubble parameter $H$).
All massive modes with mass $m\ge 5p M/2$ behave as oscillating
plane waves at infinity ($M y\to \infty$), which are delocalized
KK modes. There is only one light mode with $m=2p M$, but this
mode is non-normalizable and hence it cannot be localized on a de
Sitter brane.

Further, if we take
\begin{equation}
F(y)=\frac{y^2+ b y + c^2}{L^2},
\end{equation}
where $c$ is a resolution parameter, then the zero-mode
wavefunction is given by
\begin{equation}
\Psi\Z{0}(y) \propto \frac{F^{-3/4}}{\sqrt{{dF}/{dy}}}=
\frac{L^5\,(y^2+ b y + c^2)^{-3/4} }{(b+2y)^{1/2}}.
\end{equation}
This is normalizable, provided that $-b/2 < y <\infty$. For
brevity, we may set $b=0$ and hence restrict the $y$ coordinate in
the range $0 \lesssim \epsilon < y<\infty$.

In summary, we have found a class of physically viable models of
accelerating universe for which the 4D Newton constant is finite
and the 4D massless graviton wavefunction is normalizable. This is
quite remarkable result and it is the first of such kind in the
literature that gives an explicit realisation of 4D Einstein
gravity. The results above can easily be generalized in higher
dimensions, including 10- and 11-dimensional supergravity, with or
without a bulk cosmological term.

\subsection{Tensor perturbation equation with $\Lambda\Z{b}< 0$}

With $|\Lambda\Z{b}|\ne 0$, the perturbation equation can in
principle be analyzed by making a generic choice of the warp
factor. Here we shall simplify the model by taking
\begin{equation}
F(y) \equiv e^{My}, \quad G(y)=\frac{15 M^2 F(y)}{36 H^2
F(y)-4\Lambda\Z{b}} \quad (n=3)
\end{equation}
and restricting the $y$ coordinate in the range $0\le y < \infty$.
By defining
\begin{equation}
\psi_m(y) \equiv \frac{e^{3My/2}}{(9H^2 F-
\Lambda\Z{b})^{1/4}}\,\Psi\Z{m}(y),
\end{equation}
we can bring Eq.~(\ref{TT-tensor}) in the standard Schr\"odinger
equation (\ref{Schro-main1}), with
\begin{equation}
V(y)=\frac{M^2}{16} \frac{\left(2025 H^4 F^2 -576 F \Lambda\Z{b} +
36 \Lambda\Z{b}^2\right)}{(9 H^2 F-  \Lambda\Z{b})^2}.
\end{equation}
In the limit $My\gg 1$, $V(y) \to \frac{25 M^2}{16}$. The mass gap
now depends on the value of $M$ alone.

The massless zero-mode wavefunction, which is given by
\begin{equation}
\Psi\Z{0}(y)= N\Z{0}\, e^{-3 My/2} \left( 9 H^2 e^{My}
-\Lambda\Z{b}\right)^{1/4},
\end{equation}
where $N\Z{0}$ is the normalization constant, is normalizable. In
the $\Lambda\Z{b}=0$ case, we have
\begin{equation}
\int |\Psi\Z{0}|^2 = N\Z{0}^2\, \frac{6 H}{5M} \left[ 1-
e^{-\frac{5}{2} M y\Z{\infty}}\right].
\end{equation}
In the the $\Lambda\Z{b}< 0$ case, we define ${\cal B}\equiv
\left(-\Lambda\Z{b}\right)/9 H^2$. We then find
\begin{eqnarray}
\int_0^\infty |\Psi\Z{0}|^2 
&\simeq & \frac{N\Z{0}^2\,H}{8 M c^{2} } \Big[ (2{\cal B}-1)
\sqrt{{\cal B}+1} (4{\cal B}+3)+\frac{3}{\sqrt{{\cal B}}}\ln
\left( \sqrt{{\cal B}+1}+\sqrt{{\cal B}}\,\right)\Big].
\end{eqnarray}
In the limit ${\cal B}\to 0$, this yields
$$
\int |\Psi\Z{0}|^2 = \frac{N\Z{0}^2 H}{8M} \left(\frac{48}{5}+
\frac{24{\cal B}}{7} -\frac{2{\cal B}^2}{3} +{\cal O}({\cal
B}^3)\right) \simeq \frac{6 N\Z{0}^2 H}{5M}.$$ With ${\cal B}\gg
1$, we have $\int |\Psi\Z{0}|^2\simeq N\Z{0}^2 H \sqrt{{\cal B}}/M
= N\Z{0}^2 (-\Lambda\Z{b})^{1/2}/(3M)$. The amplitude of zero-mode
wavefunction is suppressed when the mass-like parameter $M$ is
large.

The discrete KK modes are given in terms of the Legendre
functions, i.e.,
\begin{equation}
\Psi\Z{m} = N\Z{m} \sqrt{3H} \left(e^{My} +c \right)^{1/4} \left[
P_\nu^n (z) + c\Z{1} Q_\nu^n (z)\right].
\end{equation}
where $N\Z{m}$ is the normalization constant and
$$ n\equiv \sqrt{\frac{25}{4}-\frac{4m^2}{M^2}}-\frac{1}{2}, \quad
\nu \equiv \sqrt{9-\frac{4m^2}{M^2}}, \quad z \equiv
\sqrt{1+\frac{e^{My}}{{\cal B}}}.$$ For $n$, $\nu$ integers and
$z$ real, the Legendre function of the first kind simplifies to a
polynomial, called the Legendre polynomial. This is the case only
with $m=0$.

All heavy modes with $m> 3M/2$ become oscillating plane waves,
which represent the de-localized KK massive gravitons. The
time-evolution of the mode functions of these heavy modes shows
that they remain underdamped at late times ($t\to \infty$).

\section{Conclusion}

In the standard Kaluza-Klein approach to string theory
compactification, one aims to connect the observed
four-dimensional world with the 10-dimensional physics that arises
naturally in string theory, by assuming that six of the
ten-dimensions are very tiny and compact. In this context, there
are no-go theorems, as discussed in~\cite{Malda-Nunez,GKP},
setting the constraints that a given background has to satisfy in
order to allow for solutions where the 4D spacetime is Minkowski
or de Sitter. If fluxes are present in the solution, then some
negative sources, like orientifold planes, must be present, or one
should allow for metric singularities. That means, the problem of
finding four-dimensional de Sitter solutions is well posed, if all
extra dimensions are physically compact.

In string theory, the extra dimensions of spacetime are
conjectured to take the form of a 6-dimensional Calabi–Yau
manifold. Calabi-Yau manifolds are sometimes defined as compact
K\"ahler manifolds whose canonical bundle is trivial. However, in
general, Calabi-Yau spaces are noncompact and they are also known
to allow at least one noncompact direction, admitting a warped
throat geometry or a conifold type singularity.

One can study higher-dimensional supergravity theories by allowing
a noncompact dimension. This is a perfectly viable option. In this
paper we have addressed the problem of finding de Sitter solutions
in the usual four dimensions, by studying a general class of
warped compactifications that allows one of the extra spaces to be
large in size. We have found explicit cosmological solutions of
the $(4+n)$-dimensional Einstein equations with and without a bulk
cosmological term, which could lead to a finite 4D Newton constant
as well as a massless 4D graviton localized in an inflating
four-dimensional FLRW universe. A remarkable feature of the
solutions presented in this paper is also that the size of the
radial dimension takes a constant value in the large volume limit,
providing an explicit example of spontaneous compactification. Our
examples look similar to that in Randall-Sundrum braneworld
models, but they are presented in a cosmological context.

\medskip
\noindent {\bf Acknowledgements:} I would like to thank Hideo
Kodama, Kazuya Koyama, Kei-ichi Maeda, Nobu Ohta, Shinji
Mukohyama, Valeri Rubakov and Takahiro Tanaka for valuable
conversations on topics related to this work. I would also like to
thank the YITP Kyoto University for hospitality where part of the
work was carried out. The present work is supported by the Marsden
fund of the Royal Society of New Zealand.


\end{document}